\documentclass[prb,twocolumn, groupaddress , amsmath, amssymb, showpacs, showkeys,nobibnotes,reprint,aps,longbibliography]{revtex4-1}
\usepackage{graphicx}
\usepackage{dcolumn}
\usepackage{bm}
\usepackage{upgreek}
\usepackage{times}
\usepackage{mathrsfs}
\usepackage{multirow}
\usepackage{fancyhdr} 
\usepackage{hyperref} 
\usepackage[version=3]{mhchem}

\begin{document}

\title{Localization of electronic states in III-V semiconductor alloys: a comparative study}

\author{C.~Pashartis}
\author{O.~Rubel}
\email[E-mail:~]{rubelo@mcmaster.ca}
\affiliation{Department of Materials Science and Engineering, McMaster University, 1280 Main Street West,
Hamilton, Ontario L8S 4L8, Canada}

\date{\today}

\begin{abstract}
Electronic properties of III-V semiconductor alloys are examined using first principles with the focus on the spatial localization of electronic states. We compare localization at the band edges due to various isovalent impurities in a host GaAs including its impact on the photoluminescence line widths and carrier mobilities. The extremity of localization at the band edges is correlated with the ability of individual elements to change the band gap and the relative band alignment. Additionally, the formation energies of substitutional defects are calculated and linked to challenges associated with the growth and formability of alloys. A spectrally-resolved inverse participation ratio is used to map localization in prospective GaAs-based materials alloyed with B, N, In, Sb, and Bi for 1.55~$\upmu$m wavelength telecommunication lasers. This analysis is complemented by a band unfolding of the electronic structure and discussion of implications of localization on the optical gain and Auger losses. Correspondence with experimental data on broadening of the photoluminescence spectrum and charge carrier mobilities show that the localization characteristics can serve as a guideline for engineering of semiconductor alloys.
\end{abstract}


\maketitle

%
%
\section{Introduction}\label{Sec:Introduction}

Semiconductors alloys are widely used as an active material in a variety of optoelectronic applications, such as lasers, solar cells, light-emitting diodes (LED's), and photodetectors \cite{Yamaguchi_SE_79_2005}. For instance, dilute bismides Ga(AsBi) \cite{Gogineni_APL_103_2013,Marko_SR_6_2016}, dilute nitrides (GaIn)(NAs) \cite{Broderick_SST_27_2012, Kondow_JJAP_35_1996}, and group-III nitrides (AlGaIn)N  \cite{Kneissl_SST_26_2010} represent group-III-V materials that are actively studied. Mixing of semiconductors enables tuning of their optical properties, lattice parameters (epitaxial strain), and transport characteristics to a desired application. Examples include 1.55~$\upmu$m emission wavelength in the telecommunications industry, the visible spectrum for general lighting, an ultraviolet range of wavelengths for disinfecting water, or a multi-junction monolithic solar cell with 1~eV absorption edge.

Properties of the solid solutions often deviate from Vegard's law, i.e., cannot be represented as a linear interpolation between properties of the constituents. Alloying elements perturb the electronic structure of the host material, that may introduce traps, which affect optical and transport characteristics. For instance, a band gap bowing \cite{Alberi_PRB_75_2007} and a broadening of the photoluminescence (PL) line width in dilute nitride semiconductors \cite{Rubel_JAP_98_2005}, or drastic decrease of the hole mobility in dilute bismides \cite{Tiedje_IJNT_5_2008} are attributed to band tails, which originate from the spatially localized states created by isoelectronic substitutional impurities of individual N or Bi as well as their clusters in GaAs host \cite{Kent_PRB_64_2001,Zhang_PRB_71_2005,Zhang_PRB_74_2006,Mohmad_PSSB_251_2014}. Extrinsic factors, such as non-intentional dopant impurities, native defects, and heterostructure interfaces, also contribute to localization, however we will examine intrinsic factors only in our study.

Localization of electronic states can detrimentally affect performance of devices, which should be taken into account when selecting alloying elements. In the case of solar cells, high mobility is desired for the efficient transport of photogenerated carriers to electrodes. The presence of band tails leads to inhomogeneous broadening of the gain spectrum causing reduction of the peak value of intensity gain (Ref.~\onlinecite{chow2013semiconductor}, pp.221--223). On the other hand, a wide photoluminescence line width can be a desired attribute in the case of ultraviolet light sources for water purification \cite{Kneissl_SST_26_2010} or a general-purpose LED lighting \cite{Schubert_S_308_2005}.

Electronic structure calculations are widely used to explore parameter space \cite{Dudiy_PRL_97_2006} and aid in quantifying the disorder introduced into a system by isoelectronic substitutional impurities. For example, GaP:N, GaAs:Bi, and GaN:As exhibit strong spatial localization of wave functions associated with impurity states near to the band edges \cite{Bellaiche_PRB_54_1996,Kent_PRB_64_2001,Klar_PRB_64_2001,Usman_PRB_87_2013,Bannow_PRB_93_2016}. Examining of a three-dimensional charge density distribution associated with an electronic state in an alloy provides some information about the extent of localization, but lacks a quantitative criterion that distinguishes between localized and delocalized states. 

\citet{Bellaiche_PRB_54_1996} originally proposed a localization parameter as a sum of probabilities $\rho_\alpha(E_i)$ of finding an electron in a specific eigenstate $E_i$ in the vicinity of atomic sites $\alpha$ of the same element type. \citet{Deng_PRB_82_2010} adapted this approach to a single impurity, e.g., GaAs:\textit{X}, and defined a localization ratio as the ratio of the wave function probability amplitude at the impurity site $X$ relative to the element it is substituted for in the host lattice
\begin{equation}\label{Eq:BandEdgeLocal}
\zeta= 
	\frac
	{\rho_X}
	{\rho_\text{Ga/As}} 
	- 1.
\end{equation}
The stronger the localization is, the higher the ratio; the ratio of zero corresponds to no spatial perturbations of the electronic state. The magnitude of the ratio can be used to assess the relative impact of various impurities on the band edges of the host. However, generalization of this approach to an \textit{arbitrary} state in an alloy is not straightforward due to the lack of a ``reference" state.

Localization in alloys can be defined as a vanishing zero-temperature d.c. conductivity in the system (the absence of diffusion in the \citet{Anderson_PR_109_1958} model). The value of conductivity as a function of the Fermi energy provides a spectral measure of localization. However, its calculation from first principles is a cumbersome process. Instead, \citet{Wegner_ZPB_36_1980} suggested the second moment of the wave function probability amplitude, i.e., the inverse participation ratio (IPR)
\begin{equation}\label{Eq:PR-general}
	\chi(E_i) = \frac{\int |\psi_i(\bm{r})|^4~\mathrm{d}\bm{r}}
	{\left[\int |\psi_i(\bm{r})|^2~\mathrm{d}\bm{r} \right]^2}
\end{equation}
as a criterion to distinguish between localized and extended states. The inverse value $\chi^{-1}$ represents a volume within which the wave function $\psi_i(\bm{r})$ is confined. It can also be related to the probability of an electron to return to the same state after an infinite time interval \cite{Kramer_RPP_56_1993} and thus is ultimately linked to the transport properties.

Here we use both the localization ratio and the IPR criteria to perform a comparative study of localization effects in group-III-V semiconductors alloys. First, we characterize isovalent impurities of B, N, Al, P, In, Sb, Tl, and Bi in the host GaAs.  We will show that the localization strength of these impurities is governed by their Born effective charges.
The strength of localization also determines the ability of alloying elements to change the band gap and the relative alignment of the conduction and the valence band edges (CBE and VBE), which is important for band gap engineering. Additionally, formation enthalpies of the individual impurities are calculated and linked to challenges associated with the growth of corresponding alloys.
Finally, we compare disorder characteristics of various gain medium materials for 1.55~$\upmu$m lasers, including Ga$_{1-x}$In$_x$As, dilute Ga$_{1-x}$In$_x$As$_{1-y-z}$N$_y$Sb$_z$ and GaAs$_{1-x}$Bi$_x$ alloys as well as the hypotetical Ga$_{1-x}$B$_x$As$_{1-y}$Bi$_y$ compound alloys. The results of aforementioned calculations are related to established experimental trends in transport coefficients and PL characteristics of these alloys.

%
%
\section{Computational details}\label{Sec:Method}

\subsection{Electronic structure of binaries, impurities, and alloys}

The first-principles calculations were carried out using density functional theory \cite{Kohn_PR_140_1965} (DFT) and the linear augmented plane wave method implemented in the \texttt{Wien2k} package version~14.2 \cite{Blaha_2001}. The Brillouin zone of the binary compounds primitive cell was sampled using $8\times8\times8$ \citet{Monkhorst_PRB_13_1976} mesh. The muffin tin radii $R^\text{MT}$ where set to 1.72, 1.85, 1.90, 1.95, 2.00, 2.00, 2.10, 2.31, 2.33, and 2.33~Bohr  for N, B, P, Al, Ga, As, In, Sb, Tl, and Bi, respectively. The product $R^\text{MT}_\text{min}K_\text{max}=7$, which determines the accuracy of a plane wave expansion of the wave function, was used throughout the calculations.

The lattice constants and the band structures of binary compounds (Table~\ref{Table:B}) were calculated self-consistently using \citet{Wu_PRB_73_2006} generalized gradient approximation (GGA-WC) for the exchange correlation functional. The choice of exchange correlation functional was based on preliminary studies of the band structure of GaAs \cite{Bannow_PRB_93_2016}. The Tran-Blaha modified Becke-Johnson (TBmBJ) potential \cite{Tran_PRL_102_2009} was applied to overcome deficiency of the semi-local exchange correlation functional and improve accuracy for the band gaps.

\begin{table}
    \caption{Equilibrium lattice constant $a_0$, band gap $E_\text{g}$ of binary compounds in the zinc-blende structure calculated with DFT-GGA-WC-mBJ and compared with experimental values gathered in Refs.~\onlinecite{Vurgaftman_JAP_89_2001,Vurgaftman_JAP_94_2003}.}\label{Table:B}
    \begin{ruledtabular}
        \begin{tabular}{l c r c l c}
             Binary &  \multicolumn{2}{c}{DFT}  & &  \multicolumn{2}{c}{Expt.}  \\
            \cline{2-3}\cline{5-6}
             compounds & $a_0$~({\AA}) & $E_\text{g}$~(eV) & & $a_0$~({\AA}) & $E_\text{g}$~(eV) \\
            \hline
            BAs  & 4.775 & 1.66 && $\cdots$ & $\cdots$  \\
            BBi  & 5.471 & 0.39 && $\cdots$ & $\cdots$  \\
            GaN  & 4.507 & 2.96 && 4.50 & 3.30  \\
            GaAs & 5.660 & 1.53 && 5.609  & 1.52  \\
            GaSb & 6.116 & 0.72 && 6.096 & 0.81 \\
            GaBi & 6.368 & $-1.65$ & & $\cdots$ & $\cdots$ \\
            InN  & 4.995 & 0.66 && 4.98 & 0.78 \\
            InAs & 6.091 & 0.52 && 6.058 & 0.42 \\
            InSb & 6.526 & 0.17 && 6.479 & 0.24 \\
        \end{tabular}
    \end{ruledtabular}
\end{table}

\begin{table*}
    \caption{The bond statistics used to determine the lattice constant of alloys based on Vegard's law. These concentrations are determined by the number of bonds in the supercell. Deviations from a random alloy such as enhancement of a particular bond statistics due to short-range correlations are marked in brackets. The sum of the concentrations add to one.}\label{Table:A}
    \begin{ruledtabular}
        \begin{tabular}{l c c c c c c c c c c}
            Supercell composition & \multicolumn{9}{c}{Equivalent binary concentration $c_i$} & $a = \sum_i c_i a_0(i)$ \\
            \cline{2-10}
            & BAs & BBi & GaN & GaAs & GaSb & GaBi & InN & InAs & InSb & (\AA) \\
            \hline
            In$_{34}$Ga$_{30}$As$_{64}$ 				&&&& 0.469 &&&& 0.531 && 5.89 \\
            Ga$_{38}$In$_{26}$As$_{60}$N$_{2}$Sb$_{2}$ (In-N corr.)	&&&& 0.574 & 0.020 && 0.031 & 0.363 & 0.012 & 5.81\\
            Ga$_{64}$As$_{57}$Bi$_{7}$  				&&&& 0.891 && 0.109 &&&& 5.74 \\
            Ga$_{58}$B$_6$As$_{57}$Bi$_{7}$ (B-Bi corr.)  				& 0.057 & 0.036 && 0.832 && 0.075 &&&& 5.65 \\
        \end{tabular}
    \end{ruledtabular}
\end{table*}

128-atom $4\times4\times4$ supercells were built using a two-atom primitive cell basis instead of the conventional eight-atom cell as required for calculation of the effective band structure of an alloy. The k-mesh was downscaled to $2\times2\times2$. The GGA-WC self-consistent lattice constant of $a_0=5.660$~{\AA} was used for the host GaAs in the single impurity studies. The list of alloys, their bond composition, and the expectation value of the lattice parameter are provided in Table~\ref{Table:A}. In the case of alloys, the lattice constant was obtained from a linear interpolation by taking into consideration the equivalent binary concentration of bonds. The atomic positions were optimized by minimizing Hellmann-Feynman forces acting on atoms below 2~mRy/Bohr. Calculations of the Bloch spectral weight for the effective band structure were performed using \texttt{fold2Bloch} package \cite{Rubel_PRB_90_2014}. Born effective charges for impurities were computed using \texttt{BerryPI} module \cite{Ahmed_CPC_184_2013} implemented in \texttt{Wien2k}.

A localization ratio $\zeta$ defined in Eq.~(\ref{Eq:BandEdgeLocal}) was used to evaluate the extent of localization at the band edges for single impurities. An arithmetic averaging of $\zeta$'s was performed to account for degeneracies at the band edges. We found the results to be sensitive to the size of the supercell. For instance, \citet{Deng_PRB_82_2010} reports $\zeta=3.5$ at the valence band edge for a 64-atoms GaAs:Bi supercell, which can be compared to our result of $\zeta=5.1$ for a 128-atoms supercell, but different from the value of $\zeta=16$, which is obtained under identical conditions for a 432-atoms supercell. Here we constrain the size of a supercell to 128-atoms, which implies that the localization ratio may not be fully converged with respect to cell size. Nevertheless, chemical trends in a relative localization ratio $\zeta$ computed for various impurities should be valid.

The IPR $\chi$ was used as a measure of localization in alloys. It was evaluated on the basis of probabilities of finding an electron with an eigenenergy $E_i$ within the muffin tin spheres centred at atomic sites. The atomic sites, $\alpha$, and correspondingly IPR values were split into two sub-lattices: the group-III and V. The IPR was evaluated for each sub-lattice using a discrete version \cite{Murphy_PRB_83_2011} of Eq.~(\ref{Eq:PR-general})
\begin{equation}
   \chi^{\text{III/V}} (E_i) = \dfrac
   {\sum_{\alpha} \rho_\alpha^2 (E_i)}
   {\left[\sum_{\alpha} \rho_\alpha (E_i) \right]^2}~.
\end{equation}
The summation index $\alpha$ runs over all atomic sites on the group-III/V sub-lattice. Here the participation ratio $\chi^{-1}$ represents a number of atomic sites which confine the wave function $\psi_i(\bm{r})$. The lower limit of the IPR is the inverse number of atoms in the sub-lattice that corresponds to pure Bloch states. The spectrally-resolved IPR is determined as a weighted average
\begin{equation}
   \chi(E_i) = \frac
   {\chi^{\text{III}} w^{\text{III}} + \chi^{\text{V}} w^{\text{V}}}
   {w^\text{III}+w^\text{V}},
\end{equation}
where $w=\sum_{\alpha} \rho_\alpha (E_i)$ is the eigenstate-specific weight calculated for each group of atoms. It should be noted that the muffin tin spheres capture only a portion of the wavefunction (50\% or less in typical calculations reported here). The remaining part of the wavefunction is assigned to the interstitial region and not accounted in $\rho_\alpha$ values.

The IPR is less sensitive to the size of the supercell, in contrast to the localization ratio. We obtained $\chi=0.027$ for the VBE of a 128-atoms GaAs:Bi supercell vs $\chi=0.016$ in the reference 128-atoms supercell of GaAs. In the case of 432-atoms supercell, the corresponding values are $\chi=0.0095$ and 0.0046. In both cases the IPR increases by approximately a factor of two due to the disorder associated with the impurity.

\subsection{Defect formation energy}\label{Sec:Ef}

The formation energy of isoelectronic neutral defects was calculated using the expression \cite{Freysoldt_RMP_86_2014}
\begin{eqnarray}\label{Eq:Ef}
	E^\text{f}[\text{\underline{Ga}As:}X] &=& E_\text{tot}[\text{\underline{Ga}As:}X] - E_\text{tot}[\text{GaAs}] - \nonumber \\
	& & \mu[X] + \mu[\text{Ga}],
\end{eqnarray}
where $E_\text{tot}$ is the DFT total energy and $\mu$ represents the chemical potential. The notation $\text{\underline{Ga}As:}X$ implies that the element $X$ is substituted in the place of a Ga-atom in GaAs. In the calculations, the impurity was placed in a host 128-atom $4\times4\times4$ supercell, followed by relaxation of atomic positions while keeping the size of the cell constrained at the equilibrium value for GaAs.

Here we assume a bulk source of impurities for B, Al, P, In, Sb, Tl, and Bi. The DFT total energy serves as an approximation for their chemical potential
\begin{equation}\label{Eq:mu[X]}
	\mu[X] \approx E_\text{tot}[X_\text{bulk}].
\end{equation}
The nitrogen source is N$_2$ molecule, with the chemical potential approximated by
\begin{equation}\label{Eq:mu[N]}
	\mu[\text{N}] \approx \frac{1}{2}\, E_\text{tot}[\text{N}_2].
\end{equation}
By using DFT total energies for the chemical potential we neglect the zero-point energy, the vibrational energy and entropy as well as the  pressure contributions to the free energy. The chemical potential of Ga was considered within the uncertainty
\begin{equation}\label{Eq:mu[Ga]}
	E_\text{tot}[\text{GaAs}] - E_\text{tot}[\text{As}_\text{bulk}] < \mu[\text{Ga}] < E_\text{tot}[\text{Ga}_\text{bulk}],
\end{equation}
which corresponds to Ga-poor and Ga-rich growth conditions (the lower and upper bounds, respectively). The energy width of the range is determined by the formation enthalpy of the host lattice
\begin{equation}\label{Eq:dH[GaAs]}
	\Delta H_\text{f}[\text{GaAs}] \approx E_\text{tot}[\text{GaAs}] - E_\text{tot}[\text{As}_\text{bulk}] - E_\text{tot}[\text{Ga}_\text{bulk}].
\end{equation}
We can rewrite Eq.~(\ref{Eq:mu[Ga]}) using the formation enthalphy as
\begin{equation}\label{Eq:mu[Ga]_2}
	E_\text{tot}[\text{Ga}_\text{bulk}]+\Delta H_\text{f}[\text{GaAs}]  < \mu[\text{Ga}] < E_\text{tot}[\text{Ga}_\text{bulk}].
\end{equation}
It is straightforward to modify Eqs.~(\ref{Eq:mu[X]})--(\ref{Eq:mu[Ga]_2}) for  the case of group-V type of defects, $\text{Ga\underline{As}:}X$.

The DFT total energies were obtained using the Vienna \textit{ab initio} simulation program (VASP) and projector augmented-wave (PAW) potentials \cite{Kresse_PRB_54_1996,Kresse_PRB_59_1999,Blochl_PRB_50_1994}. The Perdew-Burke-Ernzerhof generalized gradient approximation \cite{Perdew_PRL_77_1996} (PBE) for the exchange-correlation functional was utilized to maximize chemical accuracy of calculations. The cutoff energy for a plane wave expansion was set at the level of 25\% higher than the value recommended in pseudopotential files ($215\!-\!500$~eV depending on the chemical composition). Our calculations yield the formation enthalpy of $\Delta H_\text{f}=-0.65$~eV per formula unit of GaAs. This result agrees reasonably well with the experimental value of $-0.74$~eV\cite{Tmar_JCG_69_1984} giving the uncertainty of $\pm0.03$~eV/atom for reaction energies obtained with DFT-PBE \cite{Hautier_PRB_85_2012}.

\subsection{Special quasi-random alloys}

The compound alloys were modeled as random unless otherwise specified in the text. Since there are multiple possible realizations of a random alloy structure, we used a special quasi-random structures (SQS) approach \cite{Zunger_PRL_65_1990}. The alloy theoretic automated toolkit (ATAT) package \cite{VandeWalle_C_42_2013} was used to distribute alloying elements within the supercell of 128~atoms using the \texttt{mcsqs} code. The alloy state is characterized by an objective function that captures deviations of atomic correlation functions from the ideal random alloy. Our objective function included pair, triplet, and quadruplet correlations with the cutoff distances of 8.1, 5.8, and 4.1~{\AA}, respectively. The selection of cutoff distances  corresponds to the fourth, second, and first nearest neighbour atoms (respectively) on the same sublattice.

A metropolis Monte Carlo algorithm was utilized to minimize the objective function with a default annealing temperature. To assist the algorithm in finding a global minimum, 10 differently seeded simulations were run until convergence and the structure with the lowest objective function was chosen for DFT calculations. Additional constraints were imposed in the SQS search algorithm to keep the supercell lattice vectors consistent with the primitive lattice and $4\times4\times4$ multiplicity.

%
%
\section{Results and discussion}\label{Sec:Results}

\subsection{Isovalent substitutional impurities in GaAs}

We begin with studying localization effects caused by isolated impurities in GaAs. The aim is to identify a single characteristic that captures chemical trends and therefore can serve as a prediction tool for mixing alloying elements into a host lattice. The electronegativity is a straightforward candidate to begin with. In the case of dilute nitrides, the localization is attributed to a large electronegativity difference between N and As\cite{Bellaiche_PRB_54_1996} (3.04 vs 2.18 according to Pauling's scale \cite{Allred_JINC_17_1961}). On the other hand, Sb and Bi behave differently in GaAs regardless of the nearly identical electronegativity values (2.05 vs 2.02).

\begin{figure}
	\includegraphics[width=0.4\textwidth]{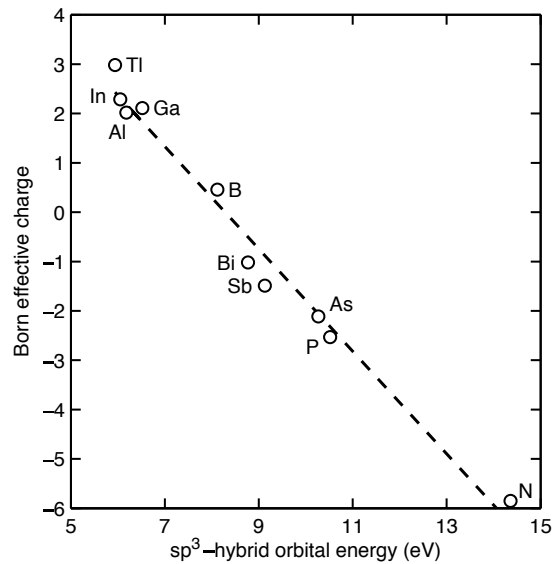}
	\caption{Born effective charge for isovalent GaAs:\textit{X} impurities as a function of their sp$^3$-hybrid orbital energy.}\label{Fig:eff_charge}
\end{figure}

The Born effective charge $Z^*$, captures the ability of an atom to attract bonding electrons by taking into consideration the particular chemical environment. The results shown in Fig.~\ref{Fig:eff_charge} indicate that Bi in GaAs is significantly more electropositive than Sb, which translates into a greater impact of Bi on the electronic structure of GaAs. From all isovalent impurities in GaAs, nitrogen shows the most striking value of $Z^*\approx-6$, which is drastically different from the value of $Z^*=-2.7$ for nitrogen in GaN \cite{Goni_PRB_64_2001}. The effective charge of Tl suggests that it is the most electropositive group-III element, contrary to the fact that both B and Tl have the identical electronegativity value\cite{Allred_JINC_17_1961}.

Calculations of the effective charge for supercells are computationally demanding. The trend line in Fig.~\ref{Fig:eff_charge} suggests that the Born effective charge scales linearly with sp$^3$-hybrid orbital energies obtained from Ref.~\onlinecite{Electron_Structure_Harrison_1989}~(pp.~50,~51), which can be used to predict the impact of an impurity on the electronic structure of the host lattice. A significance of the relative alignment of impurity-host orbital energies for the electronic stricture of alloys was previously pointed by \citet{Deng_PRB_82_2010}. The orbital energies of the valence shells together with overlap parameters are capable of capturing trap states and localization effects in various alloys, as demonstrated by O'Reilly's group from a tight-binding perspective \cite{Lindsay_PRL_93_2004,Usman_PRB_84_2011}.

\begin{figure}
	\includegraphics[width=0.49\textwidth]{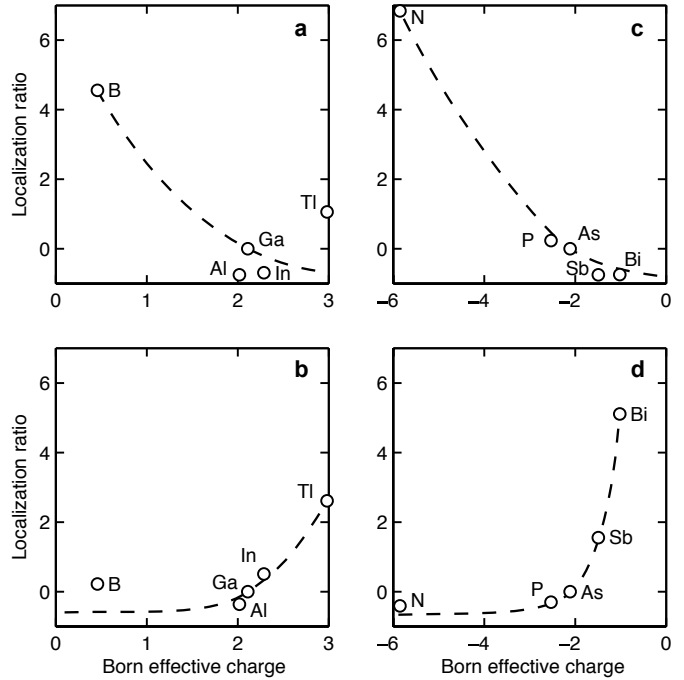}
	\caption{Localization ratio defined by Eq.~(\ref{Eq:BandEdgeLocal}) for the electronic states at the conduction (a,c) and the valence (b,d) band edges in GaAs due to single isovalent impurities plotted as a function of the element's Born effective charge. The dashed line is a guide to the eye.}\label{Fig:loc_ratio}
\end{figure}

Figure~\ref{Fig:loc_ratio} shows the localization in CBE and VBE due to isovalent impurities in GaAs. Among all impurities, nitrogen introduces the strongest localization followed by Bi and B. Those elements have more than 1$e$  difference in the Born effective charge with respect to the host elements they are substituted for. The localization ratio due to Tl and Sb is at an intermediate level. As one would expect, elements with the lower Born effective charge are effective in capturing electrons thus introducing localized states in the conduction band. The opposite applies to the valence band.

Signatures of the electron localization can be observed via broadening of the low-temperature PL line width as well as deterioration of transport coefficients. The available experimental data for ternary alloys are gathered in Table~\ref{Table:D} alongside the calculated values of the localization ratio. Based on these data, we can suggest a criterion of the localization ratio $\zeta\lesssim1$ for impurities that virtually do not impede charge carrier transport. This guideline can be used in alloy design for applications where the charge transport plays a crucial role, e.g., photovoltaics where a high mobility, bipolar charge carrier transport is desired.

\begin{table}
    \caption{Localization ratio $\zeta$ for single isovalent impurities as compared to experimental values of the charge carrier mobility $\mu$ and the PL line width  (the full width at half maximum).}\label{Table:D}
    \begin{ruledtabular}
        \begin{tabular}{l c c c r r}
            Impurity & PL line width & \multicolumn{2}{c}{$\mu/\mu_\text{GaAs}$ (\%)} & \multicolumn{2}{c}{$\zeta$} \\
            \cline{3-4}\cline{5-6}
              & (meV) & electrons & holes & CBE & VBE \\
            \hline
            
            GaAs (host) & 5\footnote{2~K \cite{Bogardus_PR_176_1968}}  & 100 \footnote{Room temperature (Ref.~\onlinecite[p. 91]{Wiley_10_1975})} & 100 \footnote{Weakly doped room temperature \cite{Stillman_JPCS_31_1970} } & $0$ & $0$ \\
            
            $\text{\underline{Ga}As:B}$ & $17.1\pm3$\footnote{1.6\% B at 300~K, pump intensity 300~W/cm$^2$, bulk-like \cite{Gottschalch_JCG_248_2003}} & $\cdots$ & $\cdots$  & $4.56$ & $0.22$\\
             
            $\text{\underline{Ga}As:Al}$ & $4, 70$\footnote{2.9\% Al, 2~K, pump intensity $10^{-3}-10$~W/cm$^2$ (Ref.~\onlinecite{Reynolds_APL_46_1985}), 3\%B at 10~K excluding substrate \cite{Hamila_JAC_506_2010}} & 96 & 96\footnote{Room temperature 1.6\%~Al (Ref.~\onlinecite[p. 626]{Shur_1990})} & $-0.75$ & $-0.36$\\
            
			$\text{\underline{Ga}As:In}$  & $2\!-\!4, 10$\footnote{Ga$_{0.46}$In$_{0.54}$As at 77~K \cite{Protzmann_JCG_170_1997}, $\approx$1\% In at 4.2K at $10^{-3}-10$~W/cm$^2$ (Ref.~\onlinecite{Mitchel_JAP_57_1985})} & $\gtrsim\!100$\footnote{Room temperature\cite{Madelung_InGaAs_2002}.} & $\cdots$  & $-0.69$ & $0.51$\\
            
            $\text{\underline{Ga}As:Tl}$ & $\cdots$ & $\cdots$ & $\cdots$  & $1.06$ & $2.61$\\
            
            $\text{Ga\underline{As}:N}$ & $30, 105$\footnote{Annealed structure at 5~K at 2\% average of 10~$\mu$W \cite{Kudrawiec_PRB_88_2013}, 4~K at 1.4\% \cite{Plaza_APL_86_2005}} & 63\footnote{Room temperature\cite{Dhar_SST_23_2007}} & $\cdots$  & $6.84$ & $-0.41$ \\
            
            $\text{Ga\underline{As}:P}$ & $ 10 $\footnote{P 30\% at 12~K quantum well structure of $\approx$ 5 quantum wells 2.7~nm thick \cite{Mitchel_JAP_57_1985}} & $\approx\!100$\footnote{Room temperature \cite{Tietjen_JES_113_1966}} & $\cdots$  & $0.23$ & $-0.30$\\
            
            $\text{Ga\underline{As}:Sb}$ & $10, 23$\footnote{Sb $\approx\!6$\% at 4.2K at (0.001-10)~W/cm$^2$ (Ref.~\onlinecite{Mitchel_JAP_57_1985}), 4.4\% Sb with excitation intensity of 2~W/cm$^2$ at 4~K \cite{Huang_JAP_63_1987}} & $\approx\!100$ & 25\footnote{8\% Sb, room temperature\cite{Bolognesi_ITED_48_2001}}  & $-0.75$ & $1.55$\\ 
            
            $\text{Ga\underline{As}:Bi}$ & $64,100$\footnote{2.6\% Bi at 10~K \cite{Oe_JJAP_41_2002}, 1.3\% at 9~K at 1~mW \cite{Francoeur_PRB_77_2008,Kini_PRB_83_2011}} & 87 & 18\footnote{Ref.~\onlinecite{Kini_PRB_83_2011}}  & $-0.74$ & $5.11$\\
        \end{tabular}
    \end{ruledtabular}
\end{table}

\begin{figure}
	\includegraphics[width=0.47\textwidth]{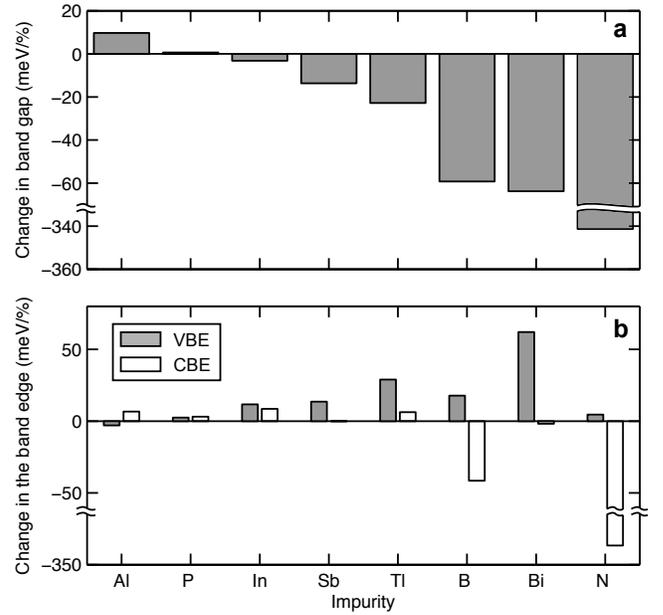}
	\caption{Effect of single impurities on the band gap (a) and the relative alignment of the band edges (b) in GaAs.}\label{Fig:induced_gap}
\end{figure}

The ability of individual element to change the band gap also scales with the localization ratio. Figure~\ref{Fig:induced_gap}(a) shows the change of a band gap of GaAs due to the single impurity, which correspond to an effective concentration of 1.6\% in the 128-atom supercell with a TBmBJ \cite{Tran_PRL_102_2009} potential. Nitrogen is by far the most effective alloying element from the band gap reduction point of view, followed by Bi, B, and sb. The band gap reduction occurs by either lowering the conduction band or shifting up the valence band edge, depending on which band edge is primarily affected by the disorder (Fig.~\ref{Fig:induced_gap},b). It is the conduction band for N and B, and conversely the valence band for Bi and Tl (see Table~\ref{Table:D}). These general results are consistent with known trends for dilute nitrides \cite{Kent_PRB_64_2001} and bismides \cite{Polak_SST_30_2015}. This information is relevant for the band gap engineering in semiconductor heterostructures, where the quantum carrier confinement is a design parameter.

\begin{figure}
	\includegraphics[width=0.45\textwidth]{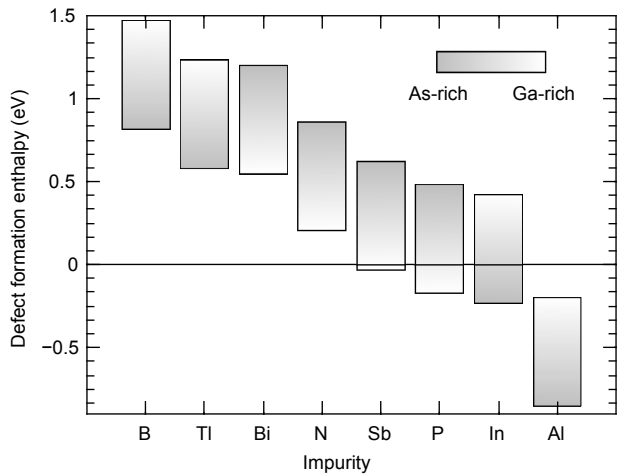}
	\caption{Formation enthalpy of isovalent substitutional defects in a GaAs host lattice. The lower the enthalpy, the more preferred the impurity is in the host system. The range of energies is linked to the growth conditions that are encoded into the gradient fill.}\label{Fig:formation}
\end{figure}

The practical implementation of compound semiconductor alloys is associated with challenges related to their formability. Growth of metastable alloys is a complex process that is largely driven by kinetic effects at the surface, rather than by thermodynamic potentials of the bulk material. A notable example is the incorporation of Bi or N in GaAs that is possible to achieve experimentally at a level, which exceeds their predicted solubility by several orders of magnitude \cite{Zhang_PRL_86_2001,Beyer_PCGCM_61_2015}. Although, the defect formation enthalphy presented in Fig.~\ref{Fig:formation} cannot be directly related to the solubility, it can still serve as a guideline for evaluating formability of alloys. Based on these data, one would expect the growth of dilute borides to pose more challenges than thallides or bismides. The energy penalty associated with incorporation of nitrogen in GaAs is at an intermediate level among other group-III-V elements. These results can be understood as an interplay between the chemical bond and local strain energies. The formation enthalpy of BAs is $-0.31$~eV\cite{Dumont_JCG_290_2006-I} per formula unit, as compared to $-1.62$~eV\cite{Ranade_JPCB_104_2000} for GaN. As a result, the strain energy due to the size mismatch between N and As is partly compensated by the strong Ga-N bond, which is not the case for the B-As bond. The incorporation of Tl and Bi in GaAs is hindered due to the same reason. It is instructive to note that the energy penalty associated with incorporation of a group-III-V element in GaAs (Fig.~\ref{Fig:formation}) is not always leveraged through the efficient band gap reduction or the effective quantum confinement engineering for semiconducting heterostructures (Fig.~\ref{Fig:induced_gap}).

\begin{figure*}
	\includegraphics[width=1\textwidth]{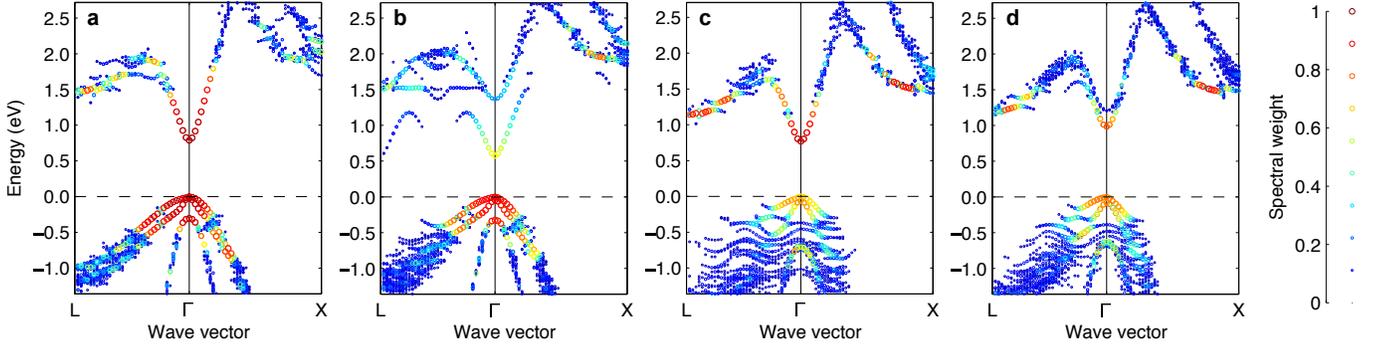}\\
	\caption{Effective band structure of semiconductor alloys for telecommunication lasers with the emission wavelength of 1.55~$\upmu$m: (a)~In$_{0.53}$Ga$_{0.47}$As, (b)~In$_{0.41}$Ga$_{0.59}$N$_{0.03}$As$_{0.94}$Sb$_{0.03}$, (c)~GaAs$_{0.89}$Bi$_{0.11}$, and (d) Ga$_{0.91}$B$_{0.09}$As$_{0.89}$Bi$_{0.11}$. The origin of the energy scale is set at the Fermi energy. The legend on the right shows the Bloch spectral weight. Only data point with the spectral weight of 5\% of greater are shown.}\label{Fig:A}
\end{figure*}

\subsection{Semiconductor alloys for telecommunication lasers}

\textbf{In$_{\bm{0.53}}$Ga$_{\bm{0.47}}$As}:~~Extending the preceding impurity study to alloys we show the link between localization characteristics, the band structure, and properties of gain materials for telecommunication lasers with the emission wavelength of 1.55~$\upmu$m. We begin with the In$_{0.53}$Ga$_{0.47}$As alloy, which represents the fist generation of gain materials.  The effective band structure is shown in Fig.~\ref{Fig:A}(a). The calculated band gap of 0.79~eV agrees well with the targeted emission wavelength. The Bloch character is well preserved in the vicinity of the band edges and persists approximately 0.5~eV into the bands. The light and heavy hole bands as well as the split-off band are clearly resolved on the unfolded band structure. The conduction band at the L-valley is energetically well isolated from the $\Gamma$-valley that prevents spurious interactions and mixing between these states in the alloy.

The above noted features of the effective alloy band structure are consistent with the narrow low-temperature PL line width ($2\!-\!4$~meV, Ref.~\onlinecite{Protzmann_JCG_170_1997}), the high mobility of electrons ($20,000\!-\!80,000$~cm$^2$V$^{-1}$s$^{-1}$, Refs.~\onlinecite{Takeda_JAP_47_1976,Protzmann_JCG_170_1997}) and holes ($600\!-\!1000$~cm$^2$V$^{-1}$s$^{-1}$, Refs.~\onlinecite{Sotoodeh_JAP_87_2000,Menon_IEE_5_2008}) observed experimentally in In$_{x}$Ga$_{1-x}$As with $x\approx0.5$. A very narrow (0.34~meV) intrinsic broadening of the band-edge alloy state in the In$_{0.5}$Ga$_{0.5}$P alloy were predicted based on empirical potential calculations \cite{Zhang_PRB_83_2011}. The low value of the localization ratio ($\zeta<1$) for GaAs:In impurities is consistent with these observations. Therefore, In$_{0.53}$Ga$_{0.47}$As can be viewed as a nearly disorder-free alloy.

\begin{figure}
	\includegraphics[width=0.49\textwidth]{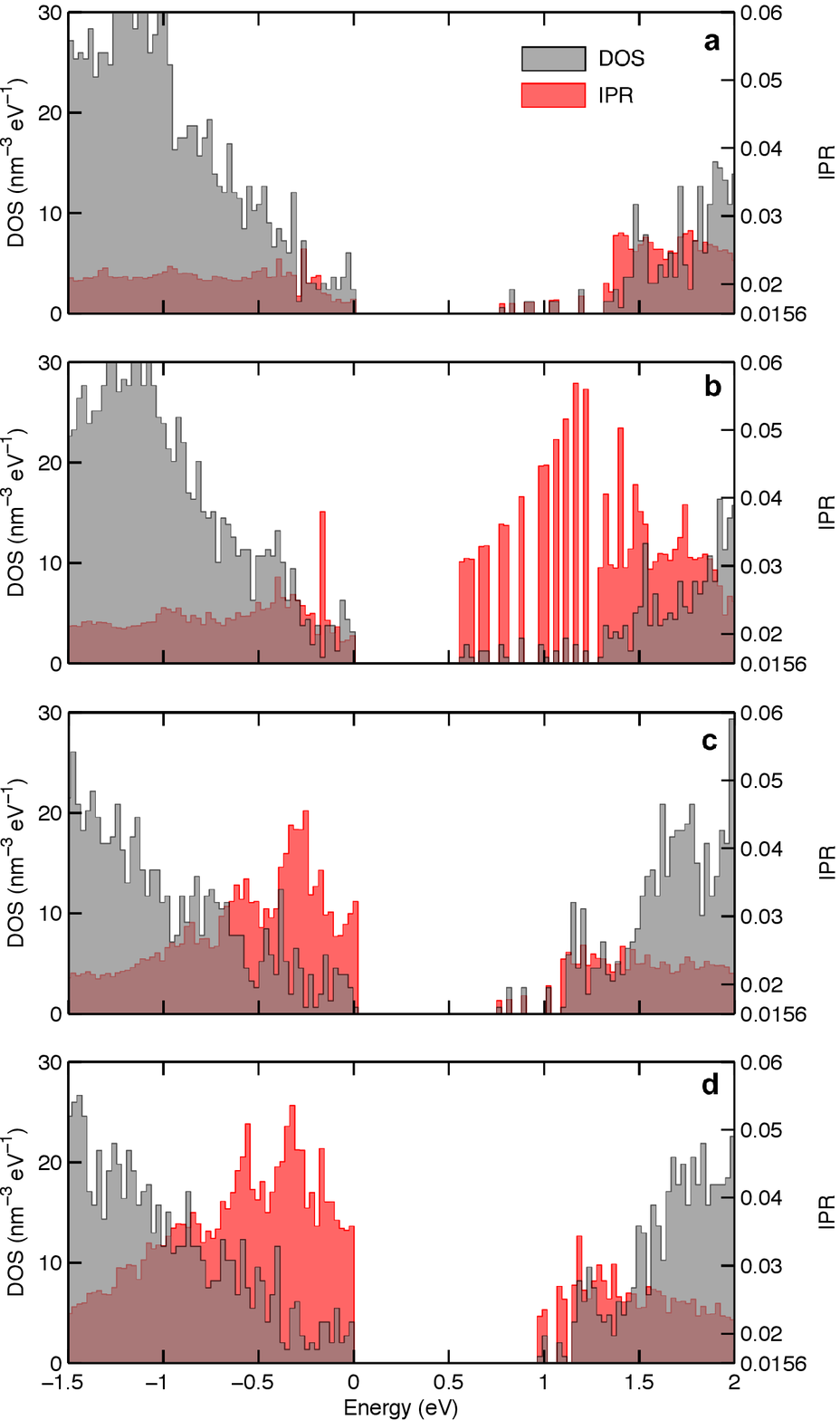}\\
	\caption{Density of states (DOS) shown alongside the inverse participation ratio (IPR), which captures the strength of localization in semiconductor alloys for telecommunication lasers with the emission wavelength of 1.55~$\upmu$m: (a)~In$_{0.53}$Ga$_{0.47}$As, (b)~In$_{0.41}$Ga$_{0.59}$N$_{0.03}$As$_{0.94}$Sb$_{0.03}$, (c)~GaAs$_{0.89}$Bi$_{0.11}$, and (d) Ga$_{0.91}$B$_{0.09}$As$_{0.89}$Bi$_{0.11}$. The lower limit of IPR 1/64 corresponds to pure Bloch states in the 128-atom supercell.} \label{Fig:E}
\end{figure}

The density of states and the IPR spectrum for the In$_{0.53}$Ga$_{0.47}$As alloy are plotted in Fig.~\ref{Fig:E}(a). The magnitude of IPR at the band edges is very close to its disorder-free limit of $\chi=1/64$, which corresponds to the 128-atom supercell. States with a higher excess kinetic energy in the valence and conduction bands show a slight increase of the IPR, in particular in the conduction band when the energy approaches the L-valley ($E\approx1.4$~eV). These states do not contribute to inhomogeneous broadening of the optical gain and do not hamper the charge carrier transport coefficients. However, states with the kinetic energy $E_\text{k}\approx E_\text{g}$ are involved in the Auger recombination, which is an important loss mechanism for lasers. Implications of the disorder for the Auger recombination will be discussed below.

\textbf{In$_{\bm{0.41}}$Ga$_{\bm{0.59}}$N$_{\bm{0.03}}$As$_{\bm{0.94}}$Sb$_{\bm{0.03}}$}:~~Dilute nitride semiconductors represent the second generation of gain materials designed for the emission wavelength of 1.55~$\upmu$m and a monolithic integration on GaAs substrate \cite{Bank_IJQE_43_2007}. The effective band structure of In$_{0.41}$Ga$_{0.59}$N$_{0.03}$As$_{0.94}$Sb$_{0.03}$ is illustrated in Fig.~\ref{Fig:A}(b). The enhanced statistics of In-N bonds is incorporated in the atomistic structure to model effect of annealing on material characteristics. The Bloch character in the valence band structure remains well-defined similar to (InGa)As, which corresponds with a similar IPR spectrum (Fig.~\ref{Fig:E},b). The CBE has drastically changed from (InGa)As. The retention of its $\Gamma$-character is only about 50\% (Fig.~\ref{Fig:A},b), which is accompanied by the abrupt increase of the IPR (i.e., localization) within the conduction band and near to the edge (Fig.~\ref{Fig:E},b). It is worth noting that the strength of localization in the conduction band \textit{increases} with increasing the electron excess energy, contrary to the common view based on the classical Anderson model of localization. The largest IPR value of nearly 0.06 can be interpreted as an orbital wave function being localized in the vicinity of $0.06^{-1}\approx17$ atoms only.  This observation is consistent with the low Bloch character (less than 30\%) for states within the energy window $E=1.0\!-\!1.3$~eV (Fig.~\ref{Fig:A},b).

These features of the electronic structure of dilute nitrides lead to broadening of the PL line width ($18\!-\!40$~meV, Refs.~\onlinecite{Rubel_JAP_98_2005,Bank_IJQE_40_2004,Goddard_JAP_97_2005,Bank_IJQE_43_2007}) and reduction of the electron mobility (100~cm$^2$V$^{-1}$s$^{-1}$, Ref.~\onlinecite{Volz_JCG_248_2003}), whereas the hole mobility remains unaffected (900~cm$^2$V$^{-1}$s$^{-1}$, Ref.~\onlinecite{Volz_JCG_248_2003}). These observations also conform with the single impurity (GaAs:N and GaAs:Sb) localization studies presented in the preceding subsection. One can anticipate a detrimental impact of localization at the CBE on the optical gain in this material system. The reason for an intrinsic ``disorder penalty" is twofold: (i) smearing of the Bloch character impedes the k-selection rules for the optical transition and reduces the dipole matrix element, (ii) the inhomogeneous broadening reduces the gain.

\textbf{GaAs$_{\bm{0.89}}$Bi$_{\bm{0.11}}$}:~~We now turn to the third generation of the gain materials, namely dilute bismide GaAs$_{1-x}$Bi$_x$ alloys. Their development is inspired by a potential ability to reduce the Auger losses \cite{Sweeney_JAP_113_2013}. The bismuth content of $x=0.11$ was selected to match the anticipated band gap with the emission wavelength of 1.55~$\upmu$m \cite{Mohmad_PSSB_251_2014}. The band structure of GaAs$_{0.89}$Bi$_{0.11}$ is shown in Fig.~\ref{Fig:A}(c). Contrary to the dilute nitrides, the bismides demonstrate a well defined profile in the conduction band, whereas the valence band is severely distorted \cite{Rubel_PRB_90_2014,Bannow_PRB_93_2016}. The conduction band minimum is well resolved with a strong $\Gamma$-character without a noticeable mixing with L- and X-states. The split-off band is located at $\Delta_\text{SO}=0.72$~eV below the valence band edge. At such a high bismuth content, the spin-orbit splitting approaches the band gap, which is $E_\text{g} =0.77$~eV in our case ($\Delta_\text{SO}>E_\text{g}$ is desired for suppression of the Auger recombination). In both material systems (InGa)As and (GaIn)(NAsSb), which were studied above, this condition is far from being fulfilled.

Examining the localization effects in GaAs$_{0.89}$Bi$_{0.11}$ (Fig.~\ref{Fig:E},c) we discern that states at the valence band edge have a degree of localization comparable to (GaIn)(NAsSb) alloys. These results explain the broad low-temperature PL line width of Ga(AsBi) alloys ($60\!-\!70$~meV, Refs.~\onlinecite{Shakfa_JAP_025709_2015,Imhof_APL_96_2010}) and the steep decline of the hole mobility ($6\!-\!15$~cm$^2$V$^{-1}$s$^{-1}$,~Refs.~\onlinecite{Nargelas_APL_98_2011,Kini_PRB_83_2011}). Dilute bismides exhibit stronger localization effects when compared to dilute nitrides, even though the localization in GaAs:Bi is less pronounced at the single impurity level than in GaAs:N ($\zeta_\text{Bi}^\text{VBE}<\zeta_\text{N}^\text{CBE}$, see Table~\ref{Table:D}). The enhanced localization in dilute bismides can be attributed to the much higher concentration of bismuth, which is required to achieve the same band gap reduction as in dilute nitrides. Accordingly, the alloy statistics leads to a higher probability of forming pairs (Bi-Ga-Bi) and higher order clusters, which have a detrimental effect on the electronic structure at the VBE \cite{Usman_PRB_84_2011,Bannow_PRB_93_2016}.

The above discussion of the ``disorder penalty" and its impact on gain characteristics is fully applicable to the Ga(AsBi) material system. The disorder also has implications on optical losses. The Auger recombination is a many-particle process, the likelihood of which is determined by the energy and momentum (wave vector) conservation of the electronic states involved. In materials with strong disorder, the momentum conservation requirement is relaxed due to uncertainties of the wave vector $\mathbf{k}$ inherent to localized states. Therefore, the inferred reduction of Auger losses in Ga(AsBi) gain media requires a thorough analysis, which will take into account the realistic band structure shown in Fig.~\ref{Fig:A}(c) and the presence of alloy disorder.

\textbf{Ga$_{\bm{0.91}}$B$_{\bm{0.09}}$As$_{\bm{0.89}}$Bi$_{\bm{0.11}}$}:~~Finally, we would like to comment of quaternary bismide alloys. For instance, Ga(NAsBi) material system \cite{Mascarenhas_SM_29_2001,Sweeney_JAP_113_2013} offers an additional flexibility in engineering of the lattice mismatch and/or band offsets.   
Here we studied a Ga$_{0.91}$B$_{0.09}$As$_{0.89}$Bi$_{0.11}$ alloy, which was inspired by the physics of In-N bonds in (GaIn)(NAs) \cite{Klar_PRB_64_2001,Kim_PRL_86_2001}. One would expect that the incorporation of boron will further reduce the band gap and help to mitigate the local strain fields by forming B-Bi bonds (a correlated alloy). In contrast, the addition of boron results in an upward shift of the $\Gamma$-valley of the conduction band and opening of the band gap (Fig.~\ref{Fig:A},d). This result suggests that the properties of quaternary alloys are governed by more complicated physics than a simple additive effects of ternary compounds. The positive effect of boron is an improvement of the Bloch character of the VBE (Fig.~\ref{Fig:A},d) in comparison to the boron-free alloy (Fig.~\ref{Fig:A},c). However, there are two drawbacks: (i) a more severe localization develops in the conduction band and deeper energy states in the valence band, (ii) the incorporation of boron in a GaAs lattice can be challenging due to the high defect formation energy (Fig.~\ref{Fig:formation}). Nevertheless, (GaB)(AsBi) could be an alternative material system for multijunction solar cells with 1~eV band gap and lattice matched to GaAs.

It is important to note that the results presented above are obtained for bulk materials, which does not take into account an epitaxial strain that can arise from a lattice mismatch between the layer and the substrate. The strain is not an issue for In$_{0.53}$Ga$_{0.47}$As and Ga$_{0.91}$B$_{0.09}$As$_{0.89}$Bi$_{0.11}$ alloys as their lattice constant is intended to match the substrates, InP and GaAs, respectively. In contrast, In$_{0.41}$Ga$_{0.59}$N$_{0.03}$As$_{0.94}$Sb$_{0.03}$ and GaAs$_{0.89}$Bi$_{0.11}$ alloys possess a lattice mismatch of 2.6 and 1.4\% with respect to GaAs (see theoretical lattice parameters in Tables~\ref{Table:B} and \ref{Table:A}). Two methods, namely \citet{Matthews_JCG_27_1974} and \citet{Voisin_PS_0861_1988}, were used to assess a critical thickness $h_\text{c}$ of the film that can accommodate the epitaxial strain elastically without generating misfit dislocations. Both approaches produce similar results of $h_\text{c}=2.5$~nm for In$_{0.41}$Ga$_{0.59}$N$_{0.03}$As$_{0.94}$Sb$_{0.03}$/GaAs and 6~nm for GaAs$_{0.89}$Bi$_{0.11}$/GaAs heterostructures. Layers of the thickness less than $h_\text{c}$ will be tetragonally distorted that may have an additional impact on localization characteristics. We evaluated the effect of a tetragonal distortion on the electronic structure and localization in the compressively-strained GaAs$_{0.89}$Bi$_{0.11}$ compound on a GaAs substrate (see Appendix). The tetragonal distortion leads to a opening of the band gap and lifting the degeneracy between heavy and light holes, which is consistent with effects expected from a compressive in-plane strain of the layer \citet{Voisin_PS_0861_1988}. There are no major changes observed in the IPR spectrum indicating that the alloy disorder is not sensitive to the epitaxial strain for the particular material system and the strain magnitude.

%
%
\section{Conclusions}\label{Sec:Conclusions}

DFT calculations were used to perform a systematic characterization of isovalent impurities (B, N, Al, P, In, Sb, Tl, and Bi) in the host GaAs. The degree of localization present in electronic states near to the band edges decreases in the following order: N, Bi, B, Tl, and Sb. Other elements do not cause notable localization of the electronic states. The localization strength scales with the Born effective charge, which was calculated for individual impurities. The effective charge of impurities can be markedly different from their nominal values in binary compounds reflecting their electropositive/electronegative nature in the specific host lattice. Elements, which have the Born effective charge less than the host, introduce localized states in the conduction band but do not perturb the valence band edge, and vice versa. The extremity of localization at the band edges directly correlates with several important characteristics, such as the charge carrier transport, broadening of the photoluminescence spectra, the ability of individual elements to change the band gap and the relative band alignment. An extensive comparison with experimental data gives confidence in the predictive power of first-principle calculations. The energy penalty associated with incorporation of impurities in the host lattice does not follow the localization pattern and decreases in the following order: B, Tl, Bi, N, and Sb. This result implies that the growth of dilute borides poses more challenges than thallides or bismides. The incorporation of nitrogen in GaAs is assisted by the strong Ga-N bond.

We performed a comparative study of localization in (InGa)As, (GaIn)(NAsSb) and Ga(AsBi) compound alloys, which represent three generations of material systems for telecommunication lasers. A wave function inverse participation ratio calculated for individual eigenstates was used as a measure for the extremity of localization. The results indicate that the electronic structure of (InGa)As is least affected by the alloy disorder. The electronic states near to the conduction band edge of (GaIn)(NAsSb) and the valence band edges of Ga(AsBi) are strongly affected by the disorder. The extremity of localization at the valence band edge of Ga(AsBi) is approximately twice greater than that for the conduction band edge of (GaIn)(NAsSb). This result translates into an intrinsically higher inhomogeneous broadening inherent to dilute bismide alloys as compared to their dilute nitride counterparts, which can have an unfavorable impact on the optical gain characteristics and Auger losses. This conclusion should not be extended to group-III antimonide semiconductors alloyed with Bi for mid-infrared devices. The smaller electronic difference between Sb and Bi should lead to a less detrimental impact of Bi on the structure of (AlGaIn)(SbBi) according to the arguments presented above for single impurities.

%
%
\begin{acknowledgments}
Authors are thankful to Axel van de Walle and Joseph Zwanziger for helpful discussions regarding generation of quasi-random structures as well as to Maciej Polak for critical reading of the manuscript. The funding is provided by the Natural Sciences and Engineering Research Council of Canada under the Discovery Grant Program RGPIN-2015-04518. The work was performed using computational resources of the Thunder Bay Regional Research Institute, Lakehead University, and Compute Canada (Calcul Quebec).
\end{acknowledgments}

%
%
\appendix*\section{Effect of the epitaxial strain (tetragonal distortion)}

\begin{figure}
	\includegraphics[width=0.47\textwidth]{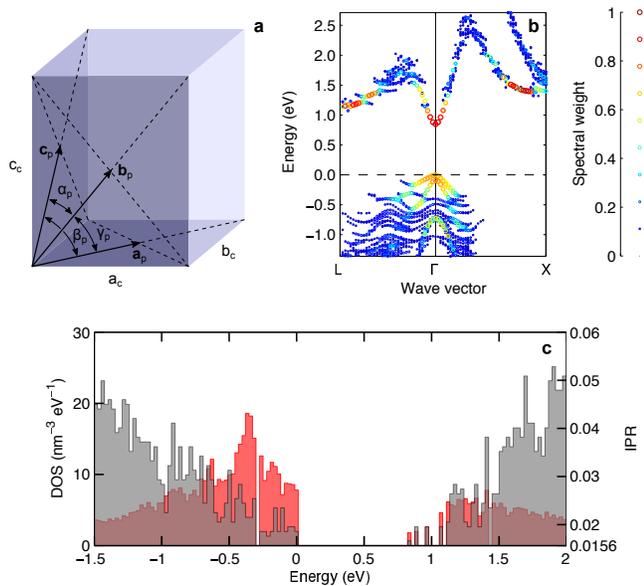}\\
	\caption{Effect of epitaxial strain on the electronic structure of GaAs$_{0.89}$Bi$_{0.11}$. (a) Conventional unit cell with a tetragonal distortion in relation to the primitive lattice vectors $\mathbf{a}_\text{p}$, $\mathbf{b}_\text{p}$, $\mathbf{c}_\text{p}$, and angles. (b) Unfolded band structure. The wave vectors are selected within the growth plane (001). (c) Density of states (DOS) and the inverse participation ratio (IPR). The origin of the energy scale is set at the Fermi energy.} \label{Fig:7}
\end{figure}

The effect of epitaxial strain on the electronic structure of GaAs$_{0.89}$Bi$_{0.11}$ was studied using a 128-atoms supercell. The lattice parameters in lateral directions ($a_\text{c}$ and $b_\text{c}$) were fixed to that of GaAs (Fig.~\ref{Fig:7},a); the $c_\text{c}$ lattice constant remained unconstrained. The lattice parameters and angles of the rhombohedral primitive cell were scaled to reproduce the tetragonal lattice distortion as shown in Fig.~\ref{Fig:7}(a).

The most noticeable changes in the electronic structure of GaAs$_{0.89}$Bi$_{0.11}$ due to the epitaxial strain are opening of the band gap by approximately 0.1~eV and lifting up the degeneracy at the top of the valence band (Fig.~\ref{Fig:7},b). The Bloch character, the density of states, and localization characteristics assessed via the inverse participation ratio do not show any prominent changes in comparison to the bulk material (Figs.~\ref{Fig:A},c and \ref{Fig:E},c). 

%
%
%

\end{document}